\documentclass[reprint,amsmath,amssymb,aps]{revtex4-1}

\usepackage{graphicx}
\usepackage{epstopdf}

\usepackage[utf8x]{inputenc}
\usepackage{hyperref}
\usepackage{color}
\usepackage[dvipsnames]{xcolor}

\usepackage{footnote}

\usepackage{subcaption}

\begin{document}
 \title{Reply to the comment on “High-Power Collective Charging of a \\Solid-State Quantum Battery” by Haowei Xu and Ju Li}

\author{Dario Ferraro$^{1,2}$}
 \email{dario.ferraro@unige.it}
 \author{Michele Campisi$^{3}$}%
\author{Gian Marcello Andolina$^{4}$}%
\author{Vittorio Pellegrini$^{5}$}%
\author{Marco Polini$^{6,7}$}%
\affiliation{%
$^{1}$ Dipartimento di Fisica,  Università  di Genova, Via Dodecaneso 33, 16146, Genova, Italy\\
$^{2}$ CNR-SPIN, Via Dodecaneso 33, 16146 Genova, Italy\\
$^{3}$NEST, Istituto Nanoscienze-CNR and Scuola Normale Superiore, I-56127 Pisa, Italy\\
$^{4}$ JEIP, UAR 3573 CNRS, Collège de France, PSL Research University,
 11 Place Marcelin Berthelot, F-75321 Paris, France\\
$^{5}$ BeDimensional Spa., 16163 Genova, Italy\\
$^{6}$ Dipartimento di Fisica dell’Università di Pisa, Largo Bruno Pontecorvo 3, I-56127 Pisa, Italy\\
$^{7}$ ICFO-Institut de Ciències Fotòniques, The Barcelona Institute of Science and Technology,
 Av. Carl Friedrich Gauss 3, 08860 Castelldefels (Barcelona), Spain}

\begin{abstract}
In this short communication we reply to the comment by Xu and Li (\href{https://doi.org/10.48550/arXiv.2411.04132}{arXiv:2411.04132}) on our first work on Dicke quantum batteries (\href{https://doi.org/10.1103/PhysRevLett.120.117702}{Phys. Rev. Lett.~\textbf{120}, 117702 (2018)}).
\end{abstract}

\maketitle

Xu and Li have recently posted a comment~\cite{arxiv_comment} on Ref.~\cite{Ferraro18}, where they express criticism on two of our findings. Even though both points they raise are valid, they have already been extensively discussed in the quantum battery literature. We therefore believe that Ref.~\cite{arxiv_comment} does not add {\it any} novelty to the quantum battery research field.

The two main criticisms that have been put forward by the authors of Ref.~\cite{arxiv_comment} to our 2018 Physical Review Letters~\cite{Ferraro18} are briefly the following:
\begin{itemize}
    \item[i)] The $\sqrt{N}$ enhancement of the average charging power does not have a genuine quantum origin.
    \item[ii)] The $\sqrt{N}$ enhancement vanishes if the density $N/V$ is fixed, where $N$ is the number of qubits and $V$ is the volume of the cavity.
\end{itemize}

These two points have already been discussed at length in previous works.

Regarding point i) above, Ref.~\cite{Andolina19}, for example, compared the charging performance of a Dicke quantum battery with that of its rigorous classical analog, finding that the power scales as $\sqrt{N}$ in both cases. As a result, Ref.~\cite{Andolina19} concluded that this scaling arises from collective effects rather than purely quantum phenomena. Similarly, Ref.~\cite{Julia-Farre20} demonstrated that the $\sqrt{N}$ scaling does not result from entanglement generation, as the battery state does not need to explore highly correlated subspaces. This issue has been further discussed in subsequent articles~\cite{zhang_arxiv_2018}. 

In passing, we would also like to comment on the following statement made by the authors of Ref.~\cite{arxiv_comment}: ``Here we would like to make it clear that the $\sqrt{N}$ enhancement in Ref.~[1] (Ref.~\cite{Ferraro18} here) is purely a classical effect, not attributable to quantum entanglement or collective phenomena.'' We fully agree (and in fact it is well established in a plethora of publications) with Xu and Li in that quantum entanglement is not at play in the model studied in Ref.~\cite{Ferraro18}. On the other hand, the fact that the non-equilibrium charging dynamics of the model studied in Ref.~\cite{Ferraro18} is a consequence of {\it collective} effects is indisputable. Indeed, as in the original Dicke model~\cite{Dicke_1954}, once the photonic degrees of freedom are integrated out, one is left with a problem where qubits interact with all-to-all effective interactions mediated by photons~\cite{Gross82}. This system is genuinely many-body and its dynamics cannot be described in a single-particle fashion. This is what we call a collective effect in condensed matter physics. We therefore find the above quoted statement by the authors of Ref.~\cite{arxiv_comment} puzzling.

As for what concerns point ii), Ref.~\cite{Julia-Farre20} emphasized that when assuming a fixed density $N/V$ of qubits in a cavity of volume $V$, the $\sqrt{N}$ enhancement indeed disappears. The authors of  Ref.~\cite{Julia-Farre20} were the first to emphasize that this assumption is necessary for the Dicke Hamiltonian to have a well-defined thermodynamic limit. However, as noted by the authors of Ref.~\cite{Andolina19} (see footnote~22 in Ref.~\cite{Andolina19}), for many finite systems of laboratory interest there are cases where $N$ can vary significantly without altering the cavity volume $V$. For example, experiments with atoms in a cavity can easily reach large values of $N$ (up to $N \sim 10^3$) in a fixed volume due to the significant difference between the ``size" of an atom (of the order of micrometers) and the cavity resonator's size (centimeters). See, for example, the recent experiment on the superabsorption properties of a Dicke-like molecular system reported in Ref.~\cite{Quach22}. 
We also note that the choice of coupling normalization in the Dicke Hamiltonian that was made in Ref.~\cite{Ferraro18}, where the qubit density is not fixed, is widely adopted in circuit-QED literature~\cite{Fink09,Wang}. For example, the Hamiltonian that was used in Ref.~\cite{Fink09} (see their Eq.~(2)) employs this very same normalization, and the discrete $\sqrt{N}$ scaling of the collective dipole coupling strength, corresponding to this normalization, was experimentally observed.

In summary, both issues i) and ii) raised by the authors of the comment to our work are well known in the community and have been thoroughly discussed in the literature, also in a number of subsequent works, in the case of different systems~\cite{Crescente20, Canzio24}. For instance, Refs.~\cite{Rossini20} and~\cite{Andolina24} proposed examples of quantum battery models displaying a {\it genuine quantum advantage}, motivated by the recognition that the advantage found in the Dicke battery is not of quantum origin. This is explicitly stated in the introductions of both papers, highlighting ongoing efforts to identify truly quantum contributions to the figures of merit of quantum battery models.

In passing, we note that Ref.~\cite{Campaioli24} thoroughly addresses issues i) and ii) above and other subtle issues in great detail. Section~III.B.3 (entitled  ``The origin of the charging advantage''), for example, is entirely devoted to point i) above while Section~III.B.4 (entitled ``Charging advantage in the thermodynamic limit") is entirely devoted to point ii) above. This two-page-long comprehensive analysis, which is contained in a Colloquium published in the prestigious Reviews of Modern Physics, indicates that the community has been actively engaged in the discussion of these issues.

Concluding, the two weak points of Ref.~\cite{Ferraro18}, highlighted by Xu and Li in Ref.~\cite{arxiv_comment}, have already been discussed in great depth in the quantum battery literature. Consequently, we believe that their comment does not provide any significant new contribution to the exciting research topic of quantum batteries.


\begin{thebibliography}{50}
%
\bibitem{arxiv_comment}  
H. Xu and J. Li,  
\href{https://doi.org/10.48550/arXiv.2411.04132}{arXiv:2411.04132}.
%
\bibitem{Ferraro18} D. Ferraro, M. Campisi, G. M. Andolina, V. Pellegrini, and M. Polini,
\href{https://doi.org/10.1103/PhysRevLett.120.117702}{Phys. Rev. Lett.~\textbf{120}, 117702 (2018)}.
%
\bibitem{Andolina19} G. M. Andolina, M. Keck, A. Mari, V. Giovannetti, and M. Polini,
\href{https://journals.aps.org/prb/abstract/10.1103/PhysRevB.99.205437}{
Phys. Rev. B \textbf{99}, 205437 (2019)}.
%
\bibitem{Julia-Farre20} S. Julià-Farré, T. Salamon, A. Riera, M. N. Bera, and M. Lewenstein,
\href{https://journals.aps.org/prresearch/abstract/10.1103/PhysRevResearch.2.023113}{
Phys. Rev. Research \textbf{2}, 023113 (2020)}.
%
\bibitem{zhang_arxiv_2018}
X. Zhang and M. Blaauboer, \href{https://doi.org/10.3389/fphy.2022.1097564}{Front. Phys. \textbf{10}, 1097564 (2023)}.
%
\bibitem{Dicke_1954}
R. H. Dicke, \href{https://doi.org/10.1103/PhysRev.93.99}{
Phys. Rev.~{\bf 93}, 99 (1954)}.
%
\bibitem{Gross82} M. Gross and S. Haroche, \href{https://www.sciencedirect.com/science/article/abs/pii/0370157382901028}{Phys. Rept. \textbf{93}, 301 (1982)}.


\bibitem{Quach22}  J. Q. Quach, K. E. McGhee, L. Ganzer, D. M. Rouse, B. W. Lovett, E. M. Gauger, J. Keeling, G. Cerullo, D. G. Lidzey, and T. Virgili,
\href{https://www.science.org/doi/epdf/10.1126/sciadv.abk3160}{Science Advances \textbf{8}, eabk3160 (2022)}.
%
\bibitem{Fink09}
J.~M.~Fink, R.~Bianchetti, M.~Baur, M.~Göppl, L.~Steffen, S.~Filipp, P.~J.~Leek, A.~Blais, and A.~Wallraff, \href{https://doi.org/10.1103/PhysRevLett.103.083601}{Phys. Rev. Lett.~{\bf 103}, 083601 (2009)}.
%
\bibitem{Wang}
{Z. Wang, {\it et al.}},
\href{https://link.aps.org/doi/10.1103/PhysRevLett.124.013601}{Phys. Rev. Lett.~{\bf 124}, 013601 (2020)}.
%
\bibitem{Crescente20} A. Crescente, M. Carrega, M. Sassetti, and D. Ferraro, \href{https://journals.aps.org/prb/abstract/10.1103/PhysRevB.102.245407}{Phys. Rev. B \textbf{102}, 245407 (2020)}.
%
\bibitem{Canzio24} A. Canzio, V. Cavina, M. Polini, and V. Giovannetti, \href{https://arxiv.org/abs/2409.20475}{arXiv:2409.20475}.
%
\bibitem{Rossini20} D. Rossini, G. M. Andolina, D. Rosa, M. Carrega, and M Polini,\href{https://journals.aps.org/prl/abstract/10.1103/PhysRevLett.125.236402}{
Phys. Rev. Lett. \textbf{125}, 236402 (2020)}.
\bibitem{Andolina24} G. M. Andolina, V. Stanzione, V. Giovannetti, and M. Polini, \href{https://export.arxiv.org/pdf/2409.08627}{arXiv:2409.08627}. 
%
\bibitem{Campaioli24} F. Campaioli, S. Gherardini, J. Q. Quach, M. Polini, and G. M. Andolina, \href{https://journals.aps.org/rmp/abstract/10.1103/RevModPhys.96.031001}{Rev. Mod. Phys. \textbf{96}, 031001 (2024)}.
%
\end{thebibliography}
\end{document}